\documentclass[aps,prb,twocolumn,groupedaddress]{revtex4}

\usepackage[T1]{fontenc}
\usepackage[english]{babel}
\usepackage{epsfig}
\usepackage{graphicx}
\usepackage{color}
\usepackage{amsmath}
\usepackage{bbold}
\usepackage{float}
\usepackage{adjustbox}
\usepackage{subfigure}
\usepackage{xifthen}
\usepackage{amsbsy}

\begin{document}

\title{Charge carrier drop at the onset of pseudogap behavior \\
 in the two-dimensional Hubbard model}

\author{Pietro~M.~Bonetti}
\author{Johannes~Mitscherling}
\author{Demetrio~Vilardi}
\author{Walter~Metzner}
\affiliation{Max Planck Institute for Solid State Research, Heisenbergstrasse 1, D-70569 Stuttgart, Germany}

\date{\today}

%
%
%
%
\begin{abstract}
We show that antiferromagnetic spin-density wave order in the two-dimensional Hubbard model yields a drop of the charge carrier density as observed in recent transport measurements for cuprate superconductors in high magnetic fields upon entering the pseudogap regime. The amplitude and the (generally incommensurate) wave vector of the spin-density wave is obtained from dynamical mean-field theory (DMFT). An extrapolation of the finite temperature results to zero temperature yields an approximately linear doping dependence of the magnetic gap $\Delta(p) \propto p^*-p$ in a broad doping range below the critical doping $p^*$. The magnetic order leads to a Fermi surface reconstruction with electron and hole pockets, where electron pockets exist only in a restricted doping range below $p^*$.
DC charge transport properties are computed by combining the renormalized band structure as obtained from the DMFT with a doping-independent phenomenological scattering rate. A pronounced drop of the longitudinal conductivity and the Hall number in a narrow doping range below $p^*$ is obtained.
\end{abstract}

\maketitle

%
%

\section{Introduction}

The structure of the normal state hidden beneath the superconducting dome determines the fluctuations that govern the anomalous properties of cuprate superconductors in a wide range of their phase diagram. \cite{Broun2008}
Recently, this normal state could be accessed in a series of experiments where superconductivity was suppressed by applying extremely high magnetic fields, up to almost 100 Tesla.
High field charge transport measurements in $\rm Y Ba_2 Cu_3 O_y$ (YBCO) and several other cuprate compounds revealed a drastic reduction of the charge carrier density upon entering the pseudogap regime.\cite{Badoux2016,Laliberte2016,Collignon2017,Proust2019} In particular, the Hall number drops from $1+p$ to $p$ in a relatively narrow range of hole doping $p$ below the critical doping $p^*$ at the edge of the pseudogap regime.
Most recently, nuclear magnetic resonance (NMR) and ultrasound experiments in high magnetic fields indicated glassy antiferromagnetic order in $\rm La_{2-x} Sr_x Cu O_4$ (LSCO) at low temperatures up to the critical doping $p^*$ for pseudogap behavior. \cite{Frachet2019} By contrast, in the superconducting state forming in the absence of a strong external magnetic field, magnetic order exists only in the low doping regime. \cite{Keimer2015}

The observed drop in charge carrier density below $p^*$ indicates a phase transition associated with a Fermi-surface reconstruction. The Hall number drop is qualitatively consistent with the formation of a N\'eel state, \cite{Storey2016,Storey2017,Verret2017} spiral magnetic order, \cite{Eberlein2016,Chatterjee2017,Verret2017,Mitscherling2018} charge order, \cite{Caprara2017,Sharma2018} and nematic order. \cite{Maharaj2017} Alternatively it may be explained by strongly fluctuating states without long-range order such as fluctuating antiferromagnets \cite{Qi2010,Chatterjee2016,Scheurer2018,Morice2017} and the Yang-Rice-Zhang state, \cite{Yang2006,Storey2016,Verret2017} while it appears difficult to relate the experimental data to incommensurate collinear magnetic order. \cite{Charlebois2017}
Magnetic scenarios received considerable support from the recent discovery of magnetic order in LSCO for any doping up to $p^*$. \cite{Frachet2019}

The competition between antiferromagnetism and superconductivity in cuprates seems to be well captured by the two-dimensional Hubbard model. \cite{Scalapino2012}
Approximate solutions of the Hubbard model indicate robust magnetic order up to fairly high doping both at moderate \cite{Yamase2016} and strong coupling, \cite{Metzner1989a,Fleck1999} provided that superconductivity is suppressed. Hence, from a theoretical point of view, the recent observation of magnetic order in LSCO by Frachet et al. \cite{Frachet2019} was not totally unexpected. Theory and experiment also agree in that N\'eel order is observed only close to half-filling, while incommensurate magnetic order dominates for sizable hole-doping. Dynamical mean-field calculations suggest that the ordering wave vector is related to the Fermi surface geometry not only at weak, but also at strong coupling. \cite{Vilardi2018}

In this work we present quantitative results for magnetic order in the Hubbard model and its impact on charge transport. The magnetic order parameter is computed from dynamical mean-field theory (DMFT) in the strong coupling regime relevant for cuprates.
Within DMFT, there is no pairing instability in the repulsive Hubbard model such that ''normal'' (non-superconducting) solutions are stable even in the absence of an external magnetic field.
We compare results for two materials with distinct band structures, namely the single-layer compound LSCO and the bilayer compound YBCO. We obtain magnetic order up to 21 percent hole-doping for LSCO, while the critical doping for magnetism in YBCO is significantly smaller.
DC charge transport properties are computed by plugging the magnetic order parameter as obtained from the DMFT calculation into expressions for the longitudinal and Hall conductivities in a magnetically ordered state derived previously by two of us. \cite{Mitscherling2018}
For the transport scattering rate we assume a constant (doping-independent) value.
We obtain a sharp drop of the longitudinal and Hall conductivity below the critical doping for magnetism in qualitative agreement with the above-mentioned high-field experiments.

Note that the DMFT yields magnetic long-range order irrespective of the low dimensionality of the system not only in the ground state, but also at low finite temperatures, since it does not capture non-local fluctuations of the magnetic order parameter orientation. However, the charge carrier density relevant for the conductivities is not strongly affected by these fluctuations, \cite{Scheurer2018} so that this deficiency of the DMFT is not a major drawback here.

This paper is structured as follows. In Sec.~II we compute the magnetic gap and the reconstructed Fermi surfaces for the two-dimensional Hubbard model with band parameters as appropriate for LSCO and YBCO. The implications for the longitudinal and Hall conductivities are discussed in Sec.~III. A conclusion in Sec.~IV closes the presentation.

%
%

\section{Magnetic order and Fermi surface reconstruction}
\label{sec:magnetism}

%
%

\subsection{Model and method}
\label{sec:method}

\subsubsection{Hubbard model and band structure}

The Hubbard model \cite{Montorsi1992} describes spin-$\frac{1}{2}$ lattice fermions with a purely local interaction. In standard second quantized notation, the Hamiltonian reads
\begin{equation}
 \mathcal{H} = \sum_{j,j',\sigma} t_{jj'} c^{\dagger}_{j,\sigma} c_{j',\sigma}
 + U \sum_{j} n_{j,\uparrow} n_{j,\downarrow} ,
\end{equation}
where $j$ and $j'$ are lattice indices, and $\sigma$ ($\uparrow$ or $\downarrow$) is the spin orientation. In applications to electrons in solids the interaction is repulsive, that is, $U>0$. Shortly after the discovery of cuprate high-temperature superconductors Anderson \cite{Anderson1987} argued that the two-dimensional Hubbard model captures the most important correlations between the valence electrons moving in the copper-oxygen planes.

We model the band structure of the copper-oxygen planes by choosing a two-dimensional square lattice with hopping amplitudes $-t$, $-t'$, and $-t''$ between nearest, second-nearest, and third-nearest neighbors, respectively. Fourier transforming this hopping matrix yields the bare dispersion relation
\begin{equation} \label{eq:dispersion}
\begin{split}
 \varepsilon_{\mathbf{k}} =&
 -2t \left( \cos{k_x} + \cos{k_y} \right) -4 t' \cos{k_x} \cos{k_y} \\
 & -2t''\left( \cos{2k_x} + \cos{2k_y} \right) .
\end{split}
\end{equation}
The ratios $t'/t$ and $t''/t$ are important material dependent parameters. \cite{Andersen1995, Pavarini2001}

The bilayer compound YBCO is modeled by two square lattice planes connected by a momentum dependent nearest-neighbor interlayer hopping amplitude $t_\mathbf{k}^\perp$. This leads to a dispersion relation
\begin{equation} \label{eq:dispersionYBCO}
 \epsilon_{\mathbf{k},k_z} = \epsilon_\mathbf{k} - t_\mathbf{k}^\perp \cos k_z ,
\end{equation}
where the two possible values $0$ and $\pi$ for $k_z$ correspond to the bonding and antibonding band, respectively.

In the following we will first describe the formalism for purely two-dimensional single-layer systems, and then mention modifications for bilayer compounds.


\subsubsection{Spiral magnetic order}

The spiral order is characterized by a finite expectation value of the local spin operator
\begin{equation} \label{eq:magnetization_orig}
 \langle\mathbf{S}_j\rangle = \frac{1}{2} \sum_{\sigma, \sigma'} \left\langle
 c^\dagger_{j,\sigma} \boldsymbol{\tau}_{\sigma\sigma'} c_{j,\sigma'}
 \right\rangle = m \hat{\mathbf{n}}_j,
\end{equation}
where $\boldsymbol{\tau} = (\tau^x,\tau^y,\tau^z)$ are the Pauli matrices, $m$ is the amplitude of the on-site magnetization (the same for all sites in a spin spiral state), and $\hat{\mathbf{n}}_j$ is a unitary vector indicating the magnetization direction on site $j$ of the form
\begin{equation}
 \hat{\mathbf{n}}_j =
 \cos\left(\mathbf{Q}\cdot\mathbf{R}_j\right)\hat{\mathbf{e}}_x -
 \sin\left(\mathbf{Q}\cdot\mathbf{R}_j\right)\hat{\mathbf{e}}_y.
\end{equation}
The magnetization thus lies in the $xy$ plane and its direction on two neighbouring sites differs by an angle $\mathbf{Q}\cdot\left(\mathbf{R}_j-\mathbf{R}_{j'}\right)$, where $\mathbf{Q}$ is an (a priori) arbitrary wave vector, and $\mathbf{R}_j$ are the coordinates of the lattice sites. For $\mathbf{Q}=\left(\pi,\pi\right)$ we recover the N\'eel state.

Fourier transforming the creation and annihilation operators, one finds that the magnetization in Eq.~(\ref{eq:magnetization_orig}) is given by the expectation value in momentum space
\begin{equation} \label{eq:magnetization}
 m = \frac{1}{2} \int \frac{d^2\mathbf{k}}{\left(2\pi\right)^2} \left\langle
 c^\dagger_{\mathbf{k}+\mathbf{Q},\uparrow} c_{\mathbf{k},\downarrow} + \text{h.c.}
 \right\rangle .
\end{equation} 
For each momentum $\mathbf{k}$, the spiral order couples only two single-particle states,
namely $\left(\mathbf{k},\downarrow\right)$ and $\left(\mathbf{k+Q},\uparrow\right)$.
It is thus convenient to use the Nambu-like basis
$(c_{\mathbf{k}+\mathbf{Q},\uparrow}, c_{\mathbf{k},\downarrow})$.
In this basis, the noninteracting Green's function has the matrix form
\begin{equation}
 \mathbf{G}^{0}_{\nu,\mathbf{k}} = \left(
 \begin{array}{cc}
 i\nu+\mu-\varepsilon_{\mathbf{k}+\mathbf{Q}} & 0 \\
 0 & i\nu+\mu-\varepsilon_{\mathbf{k}}
\end{array}\right)^{-1} ,
\end{equation}
where $\nu$ is the fermionic Matsubara frequency and $\mu$ is the chemical potential.

Spiral order in the two-dimensional Hubbard model has been found in several Hartree-Fock and slave-boson mean-field studies, \cite{Fresard1991,Igoshev2010} and in functional renormalization group calculations at moderate coupling. \cite{Yamase2016}
Spiral order has also been shown to arise naturally upon doping an antiferromagnetic Mott insulator, as described by the $t$-$J$ model. \cite{Shraiman1989,Kotov2004}


\subsubsection{Dynamical mean-field equations}

To access the strongly interacting regime, we use the DMFT, which captures non-perturbative effects such as the Mott metal-insulator transition.\cite{Georges1996}
The central approximation underlying the DMFT is a local approximation for the self-energy, which is exact in the limit of infinite lattice dimensions.\cite{Metzner1989}
Under this assumption, the Hubbard model can be mapped onto an Anderson impurity model (AIM), whose propagator is related to the lattice propagator by a self-consistency
condition. \cite{Georges1992}

Spiral magnetic order in a DMFT solution of the Hubbard model has been analyzed previously for the square lattice by Fleck et al. \cite{Fleck1999}, and for the triangular lattice by Goto et al. \cite{Goto2016}. 
Using the Nambu basis appropriate for spiral order as introduced above, the self-consistency equation has the form
\begin{equation} \label{eq:self_consistency_spiral}
 \int \frac{d^2\mathbf{k}}{\left(2\pi\right)^2}
 \left[ \big( \mathbf{G}^{0}_{\nu,\mathbf{k}} \big)^{-1} -
 \mathbf{\Sigma}^{\text{dmft}}_\nu \right]^{-1} = 
 \left( \boldsymbol{\mathcal{G}}_\nu^{-1} - \mathbf{\Sigma}^{\text{dmft}}_\nu
 \right)^{-1} ,
\end{equation}
where $\mathbf{\Sigma}_\nu^{\text{dmft}}$ is the local self-energy and $\boldsymbol{\mathcal{G}}_\nu$ is the bare propagator of the AIM. 
We parametrize the self-energy as follows
\begin{equation}
 \mathbf{\Sigma}^{\rm dmft}_{\nu} = \left(
 \begin{array}{cc}
 \Sigma_\nu & \Delta_\nu \\ \Delta_\nu & \Sigma_\nu
 \end{array} \right) ,
\end{equation}
where $\Sigma_\nu$ is the normal self-energy and $\Delta_\nu$ the magnetic gap function.
The off-diagonal term $\Delta_\nu$ is non-zero if the spin SU$(2)$ symmetry is broken by spiral magnetic order. 

In Ref.~\onlinecite{Fleck1999} a perturbative impurity solver was used for the AIM.
Here we solve the AIM associated with $\boldsymbol{\mathcal{G}}_\nu$ by exact diagonalization (ED) of the following Hamiltonian~\cite{Goto2016}
\begin{equation}
 \begin{split}
 \mathcal{H}_{\text{AIM}} = & -\mu\left(n_\uparrow+n_\downarrow\right) +
 Un_\uparrow n_\downarrow+\sum_{\ell,\sigma}\varepsilon_\ell\,
 a^\dagger_{\ell,\sigma} a_{\ell,\sigma} \\
 & + \sum_{\ell,\sigma,\sigma'} \left[V_{\ell}^{\sigma\sigma'}
 c^\dagger_\sigma a_{\ell,\sigma'} + \text{h.c.} \right] ,
 \end{split}
\end{equation}
where $c_\sigma^\dagger$ ($c_\sigma$) and $a_{\ell,\sigma}^\dagger$ ($a_{\ell,\sigma}$) are the creation (annihilation) operators of the impurity and bath electrons, respectively, and $n_\sigma = c^\dagger_\sigma c_\sigma$, while $\varepsilon_\ell$ are the bath energy levels and $V^{\sigma\sigma'}_\ell$ the hybridization amplitudes. The coefficients $V^{\uparrow\downarrow}_\ell$ account for a spin flip during a bath-to-impurity (and viceversa) hopping. They are non-zero only if magnetic order occurs.

Integrating out the bath electrons, one obtains the effective bare propagator of the impurity electrons
\begin{equation}
 \boldsymbol{\mathcal{G}}^{-1}_\nu =
 (i\nu + \mu) \mathbf{1} -
 \sum_\ell \frac{ \mathbf{V}^{\dagger}_\ell \mathbf{V}_\ell}{i\nu - \varepsilon_\ell} ,
\end{equation}
where $\mathbf{1}$ is the identity matrix in spin space and $\mathbf{V}_\ell$ is a matrix in spin space with elements $V^{\sigma\sigma'}_\ell$. 

The magnetization amplitude $m$ is obtained from the off-diagonal elements of the full DMFT propagator
$\mathbf{G}_{\nu,\mathbf{k}} =
 \big[ (\mathbf{G}^{0}_{\nu,\mathbf{k}})^{-1} -
 \mathbf{\Sigma}^{\rm dmft}_{\nu} \big]^{-1}$
in the form
\begin{equation}
 m = \frac{T}{2} \sum_\nu \int \frac{d^2\mathbf{k}}{(2\pi)^2} \left(
 G_{\nu,\mathbf{k}}^{\uparrow\downarrow} +
 G_{\nu,\mathbf{k}}^{\downarrow\uparrow} \right) .
\end{equation}

For the bilayer compound YBCO we assume an antiferromagnetic alignment between the spiral states in each layer, that is $Q_z=\pi$.
The self-consistency relation (\ref{eq:self_consistency_spiral}) is then modified to
\begin{equation}
 \frac{1}{2} \sum_{k_z=0,\pi} \int \!\! \frac{d^2\mathbf{k}}{(2\pi)^2}
 \frac{1}{\big(\mathbf{G}^{0}_{\nu,\mathbf{k},k_z}\big)^{-1} \!\!
 - \mathbf{\Sigma}^{\text{dmft}}_\nu} =
 \frac{1}{\boldsymbol{\mathcal{G}}_\nu^{-1} \! - \mathbf{\Sigma}^{\text{dmft}}_\nu} ,
\end{equation}
where
\begin{equation}
 \mathbf{G}^{0}_{\nu,\mathbf{k},k_z} = \left(
 \begin{array}{cc}
 i\nu + \mu - \epsilon_{\mathbf{k}+\mathbf{Q},k_z+Q_z} & 0 \\[1mm]
 0 & i\nu + \mu - \epsilon_{\mathbf{k},k_z}
 \end{array}\right)^{-1} .
\end{equation}

As an impurity solver for the DMFT calculation we use a recent version of an ED algorithm \cite{Kitatani2018} which is suitable for reaching relatively low temperatures. We have checked that our low temperature results are consistent with results obtained via a continuous-time quantum Monte-Carlo algorithm. \cite{Parcollet2015}

%
%
%
%

\subsection{Results}
\label{sec:magnetism:results}

We use material specific hopping amplitudes which can be calculated by downfolding {\em ab initio}\/ band structures of cuprates on a one-band Hubbard Hamiltonian. \cite{Andersen1995,Pavarini2001}
For LSCO we use $t'=-0.17t$, $t''=0.05t$ and $U=8t$, and for YBCO $t'=-0.3t$, $t''=0.15t$ and $U=10t$, with $t=0.35eV$ for both compounds. For YBCO we use a bilayer model with a momentum dependent next-nearest interlayer hopping amplitude $t_\mathbf{k}^\perp = t^\perp (\cos k_x - \cos k_y)^2$ with $t^\perp = 0.15t$. All results are presented in units of $t$.

The lowest accessible temperature at which we obtain a stable numerical solution of the DMFT self-consistency equation is $T=0.027t$ for LSCO and $T=0.04t$ for YBCO parameters. In the following we include results at $T=0.04t$ also for LSCO, to disentangle material trends from temperature dependences. We obtain homogeneous solutions for any doping.
This is in contrast to Hartree-Fock theory at zero temperature with spiral magnetic order \cite{Igoshev2010} where phase separation into regions with distinct densities frequently occurs
in a broad doping regime. 


\subsubsection{Order parameter}

\begin{figure}
\includegraphics[width=0.48\textwidth]{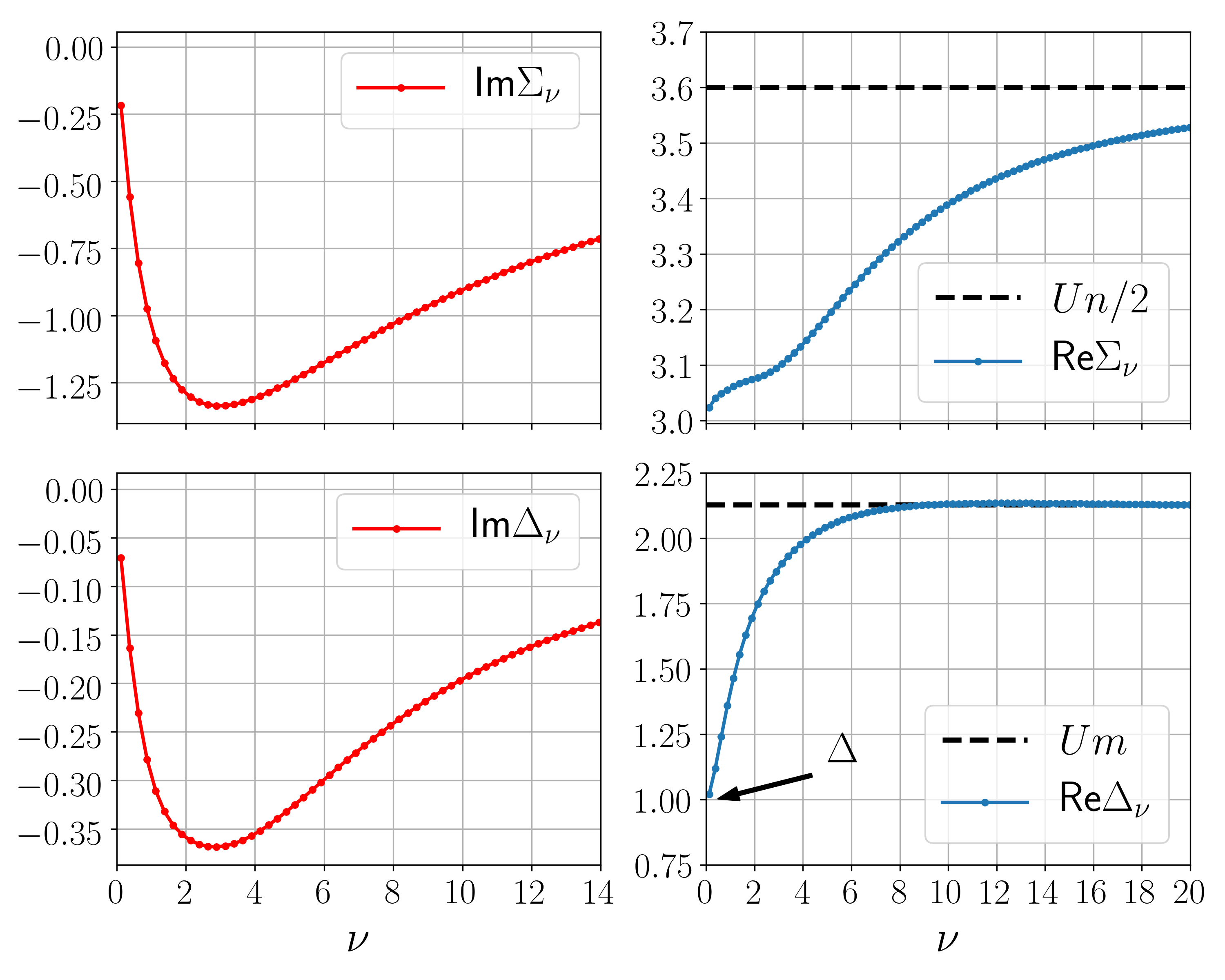}
\caption{Diagonal (top) and off-diagonal (bottom) components of the DMFT self-energy as a function of the Matsubara frequency $\nu$ for LSCO parameters at $p = 0.1$ and $T = 0.04t$.}
\label{fig:selfenergy}
\end{figure}

Previous calculations at weak coupling \cite{Schulz1990,Yamase2016} and the analysis of the DMFT magnetic susceptibility at strong coupling \cite{Vilardi2018} indicate that the 
ordering wave vector has the form $\mathbf{Q} = (\pm(\pi - 2\pi\eta),\pi)$ or $\mathbf{Q} = (\pi,\pm(\pi - 2\pi\eta))$, where $\eta \geq 0$. The value of $\eta$, usually refered to as ``incommensurability'', is determined by minimizing the DMFT free energy.\cite{Georges1996} 

In Fig.~\ref{fig:selfenergy} we show the normal self-energy $\Sigma_\nu$ and the off-diagonal self-energy $\Delta_\nu$ as a function of frequency for a specific choice of parameters in the symmetry broken phase of LSCO.
In static mean-field theory (Hartree-Fock) the off-diagonal self-energy is just a real number, yielding the magnetic gap, while in DMFT $\Delta_\nu$ is complex and frequency dependent. The frequency dependence can be sizable, but it is smooth. Its value at the lowest Matsubara frequency $\nu = \pi T$ is close to an extrapolation of $\Delta_\nu$ to $\nu = 0$.

\begin{figure}
\includegraphics[width=0.4\textwidth]{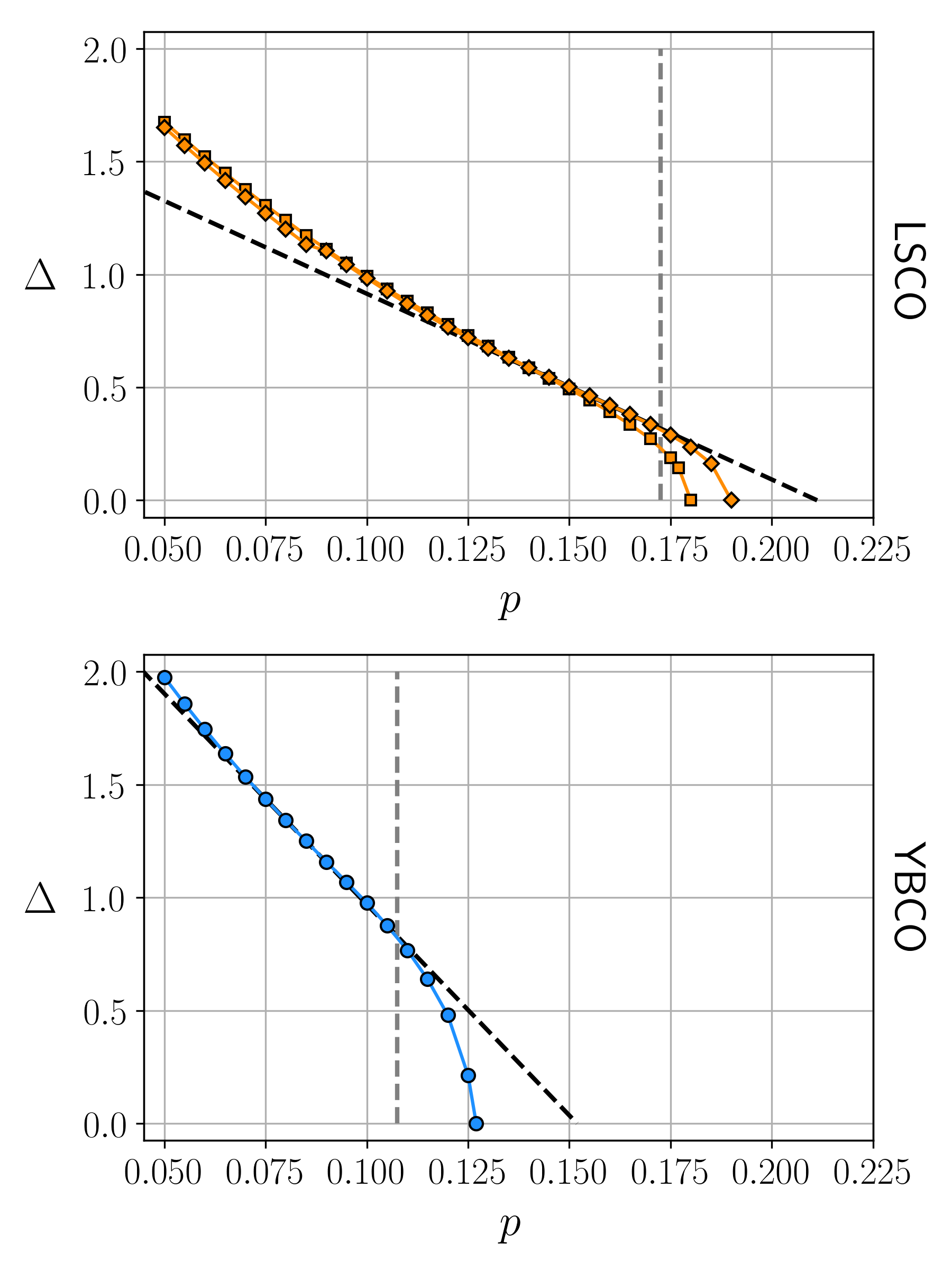}
\caption{Magnetic gap as a function of doping for LSCO at $T=0.027t$ (diamonds) and $T=0.04t$ (squares), and for YBCO at $T=0.04t$. The corresponding model parameters are described in the text. The vertical dashed line indicates the doping beyond which electron pockets are present in addition to the hole pockets (for $T=0.04t$). A linear extrapolation yielding an estimate of the gap at $T=0$ is also shown (black dashed lines).}
\label{fig:gap_vs_dop}
\end{figure}
\begin{figure}
\includegraphics[width=0.48\textwidth]{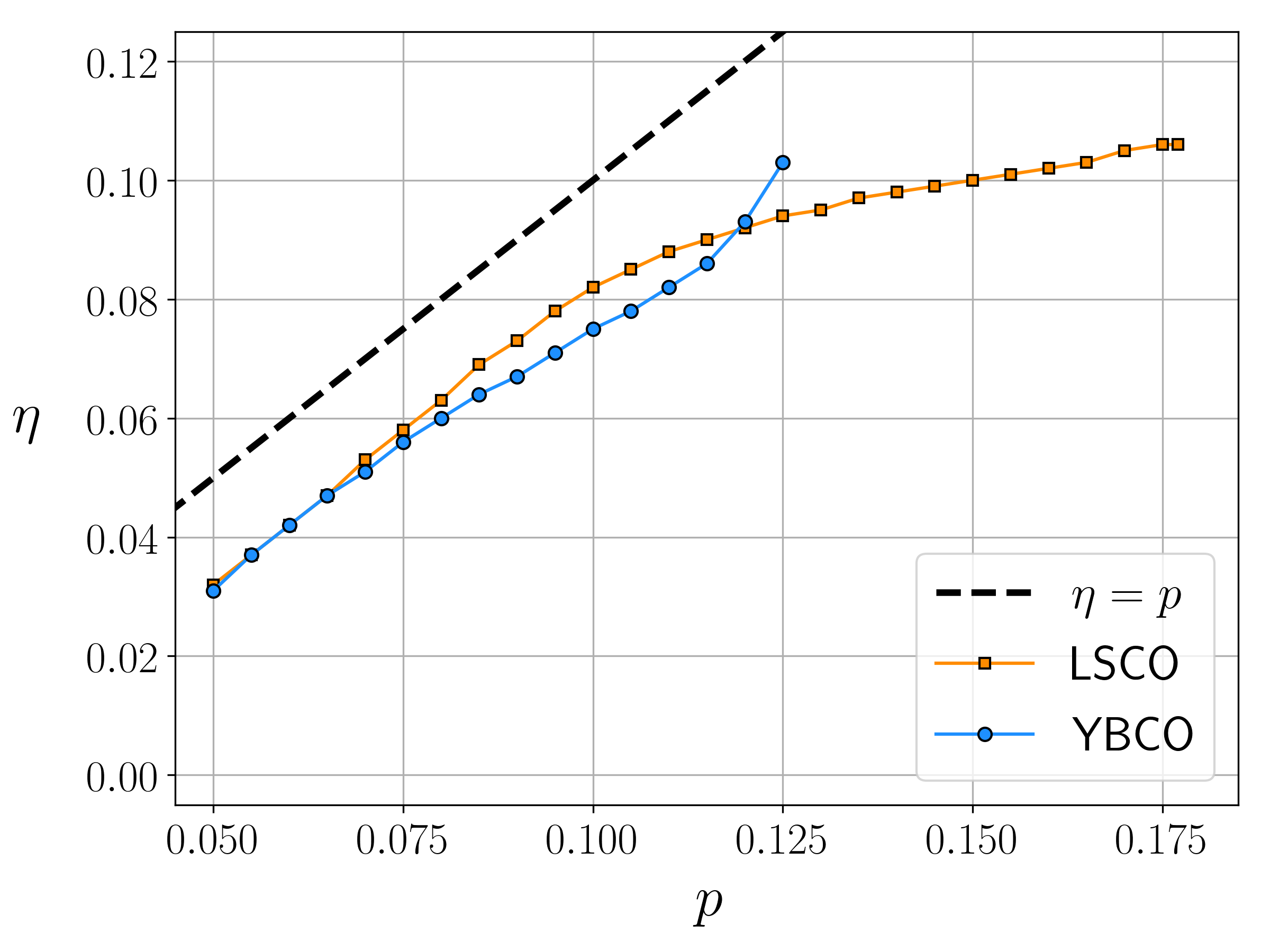}
\caption{Incommensurability $\eta$ as a function of the doping for LSCO and YBCO at $T=0.04t$.}
\label{fig:eta_vs_dop}
\end{figure}

In Fig.~\ref{fig:gap_vs_dop}, we show the extrapolated zero-frequency gap as a function of doping for LSCO and YBCO. The gap is maximal at half-filling and decreases monotonically upon doping, as expected, and vanishes continuously at the critical doping $p^*$. The magnetic order extends over a wide doping regime for all three cases. For LSCO parameters, the computed value for $p^*$ is remarkably close to the critical value for magnetic order recently observed in LSCO. \cite{Frachet2019} For YBCO parameters $p^*$ is lower than for LSCO. This is due to the larger in-plane hopping parameters beyond nearest neighbors, while the inter-plane hopping leads to a slight increase of $p^*$.
The doping range where electron pockets are present (in addition to hole pockets) is restricted to a few percent for both compounds.

The magnetic phase transition is continuous in all cases. Due to the mean-field character of the DMFT, the magnetic gap is expected to be proportional to $(p^*-p)^{1/2}$ for $p$ slightly below $p^*$ at finite temperatures, consistent with the negative curvature of $\Delta(p)$ close to $p^*$ in Fig.~\ref{fig:gap_vs_dop}. At $T=0$, which is not accessible to our calculation, $p^*$ is slightly larger, and the nearly linear $p$ dependence seen in Fig.~\ref{fig:gap_vs_dop} away from $p^*$ probably extends until the critical doping is reached, as indicated by the linear extrapolation in the figure. In principle, a weak first order transition is also possible at $T=0$.

The spiral wave vector varies with the doping, as can be seen from Fig.~\ref{fig:eta_vs_dop}, where we plot the incommensurability $\eta$ as a function of doping $p$. In all cases $\eta(p)$ is lower than $p$. Experimentally, the simple relation $\eta(p) = p$ is approximately obeyed for LSCO in the doping range $0.06 < p < 0.12$, while it saturates at $\eta \approx 1/8$ for larger doping. \cite{Yamada1998} For YBCO, the observed $\eta(p)$ is significantly lower than $p$. \cite{Haug2010}
Note that the incommensurability $\eta$ depends not only on doping, but also on temperature.
The dependence of the free energy on $\eta$ is often very weak such that the optimal choice of $\eta$ (minimizing the free energy) depends on tiny details. The impact of $\eta$ on the drop of the Hall coefficient is rather weak, \cite{Mitscherling2018} while it plays an important role in determining the degree of nematicity.


\subsubsection{Fermi surfaces}

\begin{figure}
\includegraphics[width=0.48\textwidth]{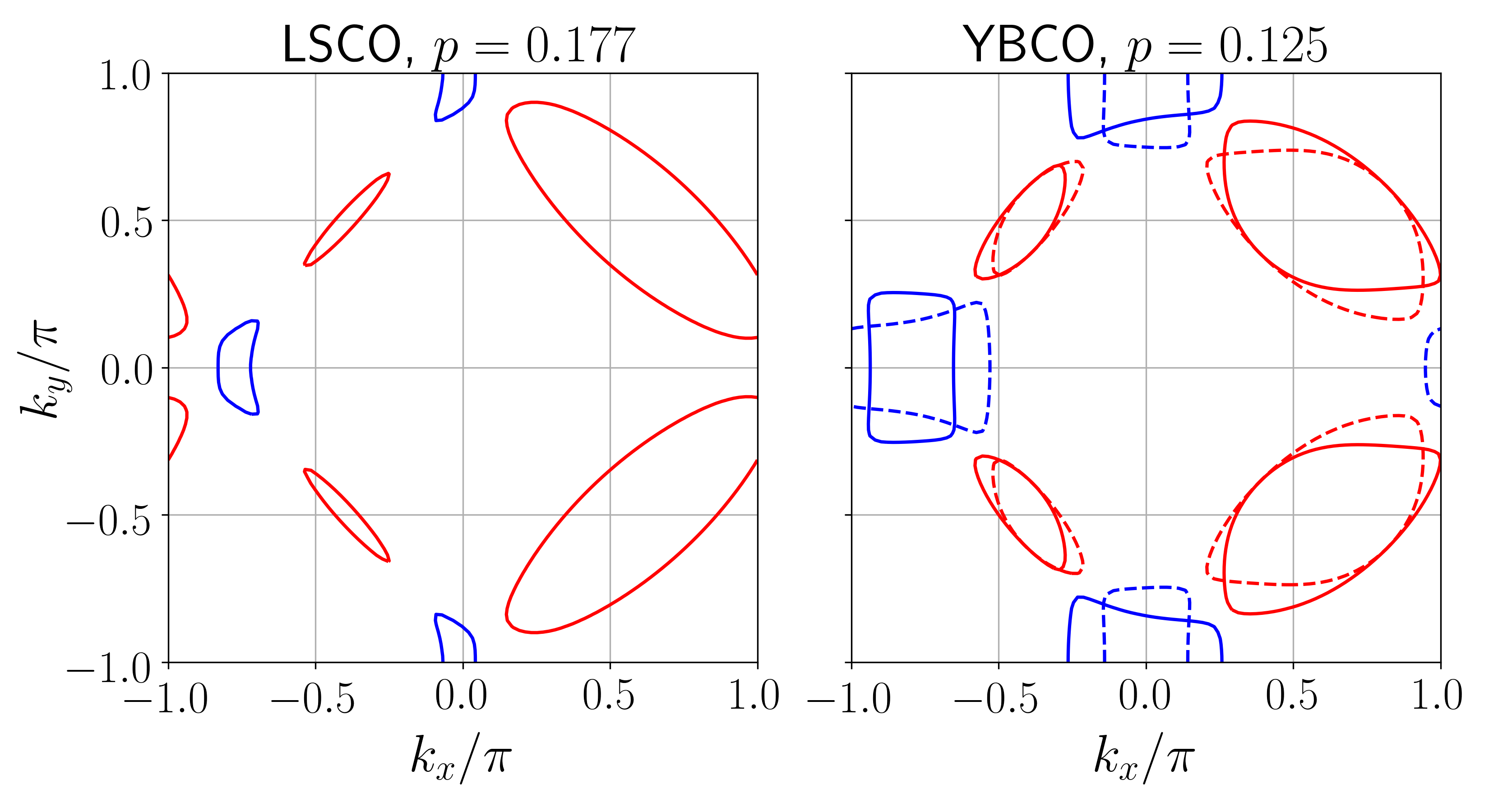}
\caption{Quasi-particle Fermi surfaces for LSCO and YBCO slightly below their respective critical doping at $T=0.04t$. Fermi surfaces of electron and hole pockets are plotted in blue and red color, respectively. For YBCO, the solid and dashed lines correspond to bonding and antibonding bands, respectively.}
\label{fig:fermi_surface_all}
\end{figure}
The magnetic gap $\Delta$ leads to a band-splitting and thus to a fractionalization of the Fermi surface. Due to the momentum independence of the DMFT self-energy, the quasi-particle bands obtained from a diagonalization of the matrix propagator for $\nu \to 0$ are given by the standard mean-field expression \cite{Igoshev2010}
\begin{equation}
 E_\mathbf{k}^\pm =
 \frac{1}{2} (\epsilon_\mathbf{k} + \epsilon_{\mathbf{k}+\mathbf{Q}}) \pm
 \sqrt{ \frac{1}{4} (\epsilon_\mathbf{k} - \epsilon_{\mathbf{k}+\mathbf{Q}})^2
 + \Delta^2} - \tilde\mu ,
\end{equation}
where $\Delta$ is the zero frequency extrapolation of $\Delta_\nu$, and
$\tilde\mu = \mu - {\rm Re}\Sigma_{\nu \to 0}$ is the renormalized chemical potential.
The quasi-particle Fermi surfaces are given by $E_\mathbf{k}^\pm = 0$.
For YBCO one has to replace $\epsilon_\mathbf{k}$ by $\epsilon_{\mathbf{k},k_z}$ with $k_z = 0,\pi$.

In Fig.~\ref{fig:fermi_surface_all} the quasi-particle Fermi surfaces of LSCO and YBCO are shown for doping values close to their respective critical doping $p^*$ on the magnetically ordered side. For YBCO, the bilayer structure splits the Fermi surface into two branches from the bonding and antibonding bands.
In all cases, both electron and hole-like pockets are present, as expected, due to the relatively small value of $\Delta$ in the vicinity of the critical doping.

The quasi-particle Fermi surface must not be confused with the Fermi surface seen in photoemission. The latter is determined by poles of the diagonal elements of the matrix Green's function at zero frequency, corresponding to peaks in the spectral function at zero frequency, $A_\mathbf{k}(0)$.
Discarding the normal self-energy except for the chemical potential shift by ${\rm Re}\Sigma_{\nu \to 0}$, and approximating $\Delta_\nu$ by its zero frequency extrapolation $\Delta$, the spectral function for spin up and spin down electrons is given by \cite{Eberlein2016}
\begin{equation} \label{eq:specfct}
 \begin{split}
 \hskip -2mm
 A_{\mathbf{k},\uparrow}(\omega) &= \sum_{n=\pm} \frac{\Delta^2}
 {\Delta^2 + (\xi_{\mathbf{k}-\mathbf{Q}} \! - \! E_{\mathbf{k}-\mathbf{Q}}^{-n})^2} \,
 \delta(\omega \! - \! E_{\mathbf{k}-\mathbf{Q}}^n) , \\
 \hskip -2mm
 A_{\mathbf{k},\downarrow}(\omega) &=
 \sum_{n=\pm} \frac{\Delta^2}{\Delta^2 + (\xi_\mathbf{k} - E_\mathbf{k}^n)^2} \,
 \delta(\omega - E_\mathbf{k}^n) ,
 \end{split}
\end{equation}
where $\xi_\mathbf{k} = \epsilon_\mathbf{k} - \tilde\mu$. For inversion symmetric dispersion relations ($\epsilon_\mathbf{k} = \epsilon_{-\mathbf{k}}$), as in our case, one can show that
$A_{\mathbf{k},\uparrow}(\omega) = A_{-\mathbf{k},\downarrow}(\omega)$.
The Fermi surface corresponding to peaks in
$A_\mathbf{k}(\omega) = A_{\mathbf{k},\uparrow}(\omega) + A_{\mathbf{k},\downarrow}(\omega)$ at $\omega=0$ is thus given by the points in momentum spaces obeying $E_\mathbf{k}^\pm = 0$ or $E_{\mathbf{k} - \mathbf{Q}}^\pm = 0$. The latter equation is equivalent to $E_{-\mathbf{k}}^\pm = 0$ for inversion symmetric $\epsilon_\mathbf{k}$.
Note that $E_\mathbf{k}^\pm$ and the quasi-particle Fermi surfaces are not inversion symmetric, while $A_\mathbf{k}(\omega)$ is.
The spectral weight on the Fermi surface is given by $\frac{\Delta^2}{\Delta^2 + \xi_\mathbf{k}^2}$, which is maximal for momenta close to the bare Fermi surface, where $\xi_\mathbf{k} = 0$.

At low energies and low temperature, the main effect of the normal self-energy $\Sigma_\nu$ is a renormalization of the quasi-particle energies and a reduction of the quasi-particle weight by the $Z$-factor
\begin{equation}
 Z = \Big[ 1 - \frac{\partial{\rm Im}\Sigma_\nu}{\partial\nu} \Big|_{\nu=0} \,
 \Big]^{-1} .
\end{equation}
At finite temperatures the differential quotient may be approximated by the quotient
${\rm Im}\Sigma_{\nu_0}/(\pi T)$, where $\nu_0 = \pi T$ is the lowest positive Matsubara frequency.
The $Z$-factor reduces the bare single-particle excitation energy and the gap to
$\bar\xi_\mathbf{k} = Z \xi_\mathbf{k}$ and $\bar\Delta = Z \Delta$, and thus also the quasi-particle energies to $\bar E_\mathbf{k}^\pm = Z E_\mathbf{k}^\pm$.
Moreover, it reduces the quasi-particle contributions Eq.~(\ref{eq:specfct}) by a global factor $Z$. The missing spectral weight is shifted to incoherent contributions at higher energies.
The spectral function for single-particle excitations can thus be written as
\begin{eqnarray}
 A_\mathbf{k}(\omega) &=&
 \sum_{n=\pm} \frac{Z \, \bar\Delta^2}{\bar\Delta^2 +
 (\bar\xi_\mathbf{k} - \bar E_\mathbf{k}^n)^2} \,
 \delta(\omega - \bar E_\mathbf{k}^n) + (\mathbf{k} \to \mathbf{-k}) \nonumber \\
 &+& A_\mathbf{k}^{\rm inc}(\omega) .
\end{eqnarray}
\begin{figure}
\includegraphics[width=0.48\textwidth]{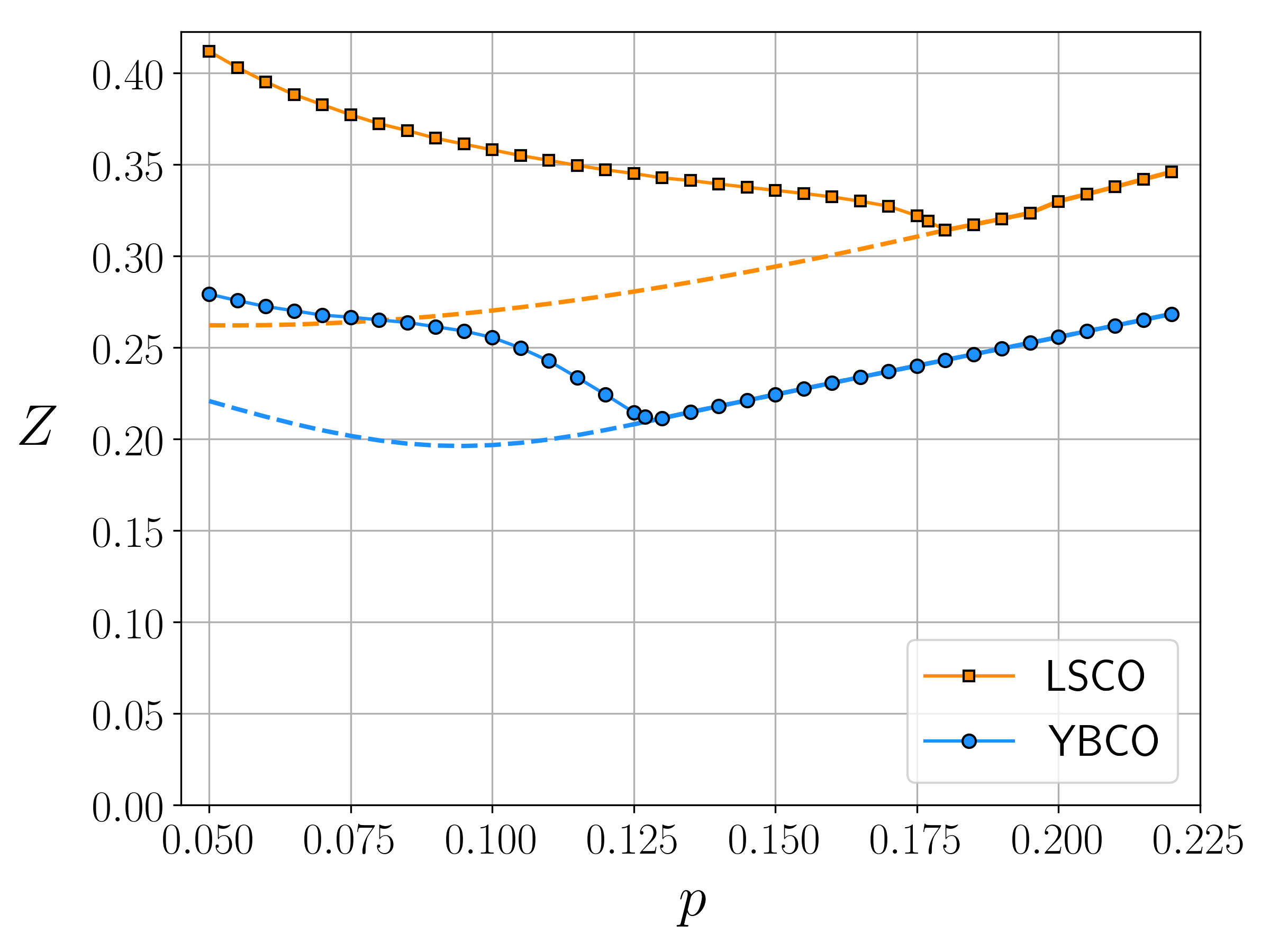}
\caption{$Z$-factor as a function of doping for LSCO and YBCO at $T=0.04t$. The $Z$-factor obtained from the unstable (below $p^*$) paramagnetic solution is also shown for comparison (dashed lines).}
\label{fig:Z}
\end{figure}
\begin{center}
\begin{figure*}
\includegraphics[width=0.9\textwidth]{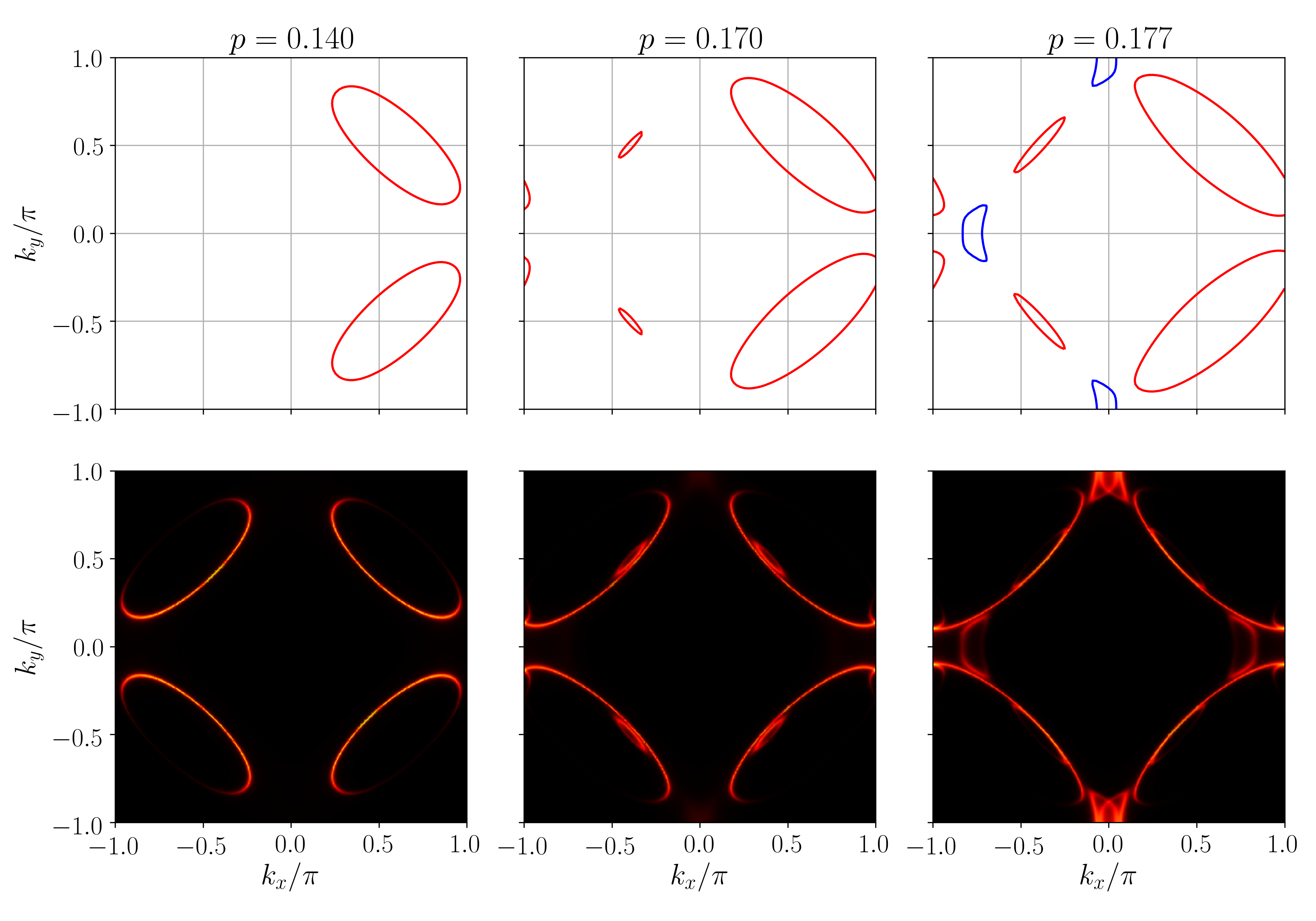}
\caption{Quasi-particle Fermi surfaces (top) and spectral functions $A_\mathbf{k}(0)$ (bottom) for LSCO band parameters at various doping values. The $\delta$-peaks in the spectral functions have been broadened by inserting a constant scattering rate $\Gamma = 0.025t$.}
\label{fig:fermi_surface_dopings}
\end{figure*}
\end{center}

In Fig.~\ref{fig:Z} we show the $Z$-factor as obtained from the DMFT as a function of doping. For $p< p^*$ we also show the $Z$-factor found in the unstable paramagnetic solution. One can see that the magnetic order enhances $Z$ compared to the paramagnetic phase. The $Z$-factor exhibits only a moderate doping dependence and assumes material-dependent values between $0.2$ and $0.4$. The strongest renormalization is found for YBCO.
Note that the paramagnetic $Z$-factors do not vanish for $p \to 0$, because for our choice of parameters the paramagnetic DMFT solution at half-filling is still on the metallic side of the Mott transition.

In Fig.~\ref{fig:fermi_surface_dopings} we plot the quasi-particle Fermi surface and the quasi-particle contribution to the spectral function $A_\mathbf{k}(0)$ in the case of LSCO for various dopings. Electron pockets are present only very close to the critical doping $p^*$. In the plots of the spectral function only the inner side of the pocket Fermi surface close to the bare Fermi surface is visible since the contributions on the outer side are suppressed by a drastically reduced spectral weight. \cite{Eberlein2016} Hence, the Fermi surface seen in photoemission seems to evolve smoothly from a large Fermi surface at high doping into Fermi arcs in the pseudogap regime. The backbending edges of the arcs could be suppressed by a more realistic momentum-dependent scattering rate.

%
%
%
%
\section{Charge transport}
\label{transport}
%
%
%
%
\subsection{Method}
\label{sec:transport:method}

In principle, one could compute charge transport properties from linear response theory within the DMFT approximation. \cite{Georges1996} However, this involves a rather delicate analytic continuation from Matsubara to real frequencies. Moreover, the scattering rates obtained from the DMFT cannot be expected to provide a good approximation in two dimensions.
Hence, we compute only the magnetic gap, the incommensurability, and the $Z$-factor from the DMFT, while we take the scattering rates from estimates obtained from experiments.

More precisely, we compute the charge conductivities for the mean-field Hamiltonian
\begin{equation} \label{eq:H_mf}
 H_{\rm MF} = \sum_{\mathbf{k},\sigma} \epsilon_\mathbf{k} 
 c_{\mathbf{k},\sigma}^\dag c_{\mathbf{k},\sigma}^{\phantom\dag} +
 \sum_\mathbf{k} \Delta \big(
 c_{\mathbf{k}+\mathbf{Q},\uparrow}^\dag c_{\mathbf{k},\downarrow}^{\phantom\dag}
 + c_{\mathbf{k},\downarrow}^\dag c_{\mathbf{k}+\mathbf{Q},\uparrow}^{\phantom\dag}
 \big) ,
\end{equation}
where the gap $\Delta$ and the ordering vector $\mathbf{Q}$ are extracted from the DMFT solution as described in the previous section. The chemical potential $\mu$ is adapted such that it corresponds to a doping $p$ in the solution of $H_{\rm MF}$.
The scattering rate is implemented by adding a small imaginary part $i\Gamma$ to the inverse retarded bare propagator.

The ordinary electrical conductivity $\sigma^{\alpha\beta}$ and the Hall conductivity $\sigma^{\alpha\beta\gamma}_H$ are defined as 
\begin{equation}
 j^\alpha =
 \left[ \sigma^{\alpha\beta} + \sigma_H^{\alpha\beta\gamma} B^\gamma \right] E^\beta \, ,
\end{equation}
where $j^\alpha$ is the current in direction $\alpha = x,y,z$. $B^\gamma$ and $E^\beta$ are the components of the external magnetic and electric field, respectively.
Building on previous work by Voruganti et al., \cite{Voruganti1992} exact expressions for the conductivities in a spiral magnetic state as described by the mean-field Hamiltonian $H_{\rm MF}$, Eq.~(\ref{eq:H_mf}), have been derived recently. \cite{Mitscherling2018} 

The longitudinal conductivity is obtained as a sum of intra- and interband contributions $\sigma^{\alpha\alpha} =
\sigma^{\alpha\alpha}_\text{intra} + \sigma^{\alpha\alpha}_\text{inter}$, with
\begin{eqnarray} \label{eq:sg_intra}
 \sigma^{\alpha\alpha}_\text{intra} \! &=& \!
 - e^2 \frac{\pi}{L} \sum_\mathbf{k} \sum_{n=\pm} \int \! d\epsilon \, f'(\epsilon) 
 E_\mathbf{k}^{n,\alpha} E_\mathbf{k}^{n,\alpha} [A_\mathbf{k}^n(\epsilon)]^2 ,
 \hskip 6mm \\
 \label{eq:sg_inter}
 \sigma^{\alpha\alpha}_\text{inter} \! &=& \!
 - 2 e^2 \frac{\pi}{L} \sum_\mathbf{k} \int \! d\epsilon \, f'(\epsilon)
 F_\mathbf{k}^\alpha F_\mathbf{k}^\alpha A_\mathbf{k}^+(\epsilon) A_\mathbf{k}^-(\epsilon) ,
\end{eqnarray}
where $L$ is the number of lattice sites, $e$ is the electron charge, and $f'(\epsilon)$ is the first derivative of the Fermi function $f(\epsilon) = \big( e^{\epsilon/T} + 1 \big)^{-1}$.
\begin{equation}
 A_\mathbf{k}^\pm(\epsilon) =
 \frac{\Gamma/\pi}{(\epsilon - E^\pm_\mathbf{k})^2 + \Gamma^2}
\end{equation}
is the spectral function of the quasiparticles, which must not be confused with the spectral function $A_\mathbf{k}(\omega)$ for single-particle excitations discussed in the preceding section. The scattering rate $\Gamma$ is assumed to be momentum independent.
$E^{\pm,\alpha}_\mathbf{k} = \partial E^\pm_\mathbf{k}/\partial k_\alpha$ are the quasi-particle velocities, and
$F^\alpha_\mathbf{k} = [\Delta/(E^+_\mathbf{k} - E^-_\mathbf{k})]
 \partial (\epsilon_{\mathbf{k}+\mathbf{Q}} - \epsilon_\mathbf{k})/\partial k_\alpha$.
Note that $\sigma^{xy} = \sigma^{yx} = 0$ as the ordering vector $\mathbf{Q}$ is N\'eel antiferromagnetic (component $\pi$) in one direction. \cite{Mitscherling2018}

The Hall conductivity is given by
$\sigma^{xyz}_H = \sigma^{xyz}_{H,\text{intra}} + \sigma^{xyz}_{H,\text{inter}}$ with
\begin{widetext}
\begin{eqnarray} \label{eq:sigmaH_intra}
 \sigma_{H,{\rm intra}}^{xyz} &=& 
 e^3 \frac{\pi^2}{3L} \sum_\mathbf{k} \sum_{n=\pm} \int d\epsilon \, f'(\epsilon)
 \big[ A_\mathbf{k}^n(\epsilon) \big]^3
 \big[ (E_\mathbf{k}^{n,x})^2 E_\mathbf{k}^{n,yy} -
 E_\mathbf{k}^{n,x} E_\mathbf{k}^{n,y} E_\mathbf{k}^{n,xy} +
 (x \leftrightarrow y) \big] , \\[2mm]
 \label{eq:sigmaH_inter}
 \sigma_{H,{\rm inter}}^{xyz} &=&
 - \, e^3 \frac{\pi^2}{L} \sum_\mathbf{k} \sum_{n=\pm} \int\! d\epsilon \,
 f'(\epsilon) \big[ A_\mathbf{k}^n(\epsilon) \big]^2 A_\mathbf{k}^{-n}(\epsilon) \,
 \big[ F^x_\mathbf{k} \big( H^{yx}_\mathbf{k} E^{n,y}_\mathbf{k} -
 H^{yy}_\mathbf{k} E^{n,x}_\mathbf{k} \big) +
 (x \leftrightarrow y) \big] \nonumber \\
 && + \, 2 e^3 \frac{\pi^2}{L} \sum_\mathbf{k} \sum_{n=\pm} \int\! d\epsilon \,
 f(\epsilon) \, A_\mathbf{k}^+(\epsilon) A_\mathbf{k}^-(\epsilon) \,
 \frac{A_\mathbf{k}^+(\epsilon) - A_\mathbf{k}^-(\epsilon)}
 {E_\mathbf{k}^+ - E_\mathbf{k}^-} \nonumber \\
 && \hspace{2cm} \times \, \left[
 F^x_\mathbf{k} \big( H^{yx}_\mathbf{k} E^{n,y}_\mathbf{k} -
 H^{yy}_\mathbf{k} E^{n,x}_\mathbf{k} +
 F^x_\mathbf{k} E^{n,yy}_\mathbf{k} - F^y_\mathbf{k} E^{n,yx}_\mathbf{k} \big) +
 (x \leftrightarrow y) \right] ,
\end{eqnarray}
\end{widetext}
where $E^{\pm,\alpha\beta}_\mathbf{k} =
\partial^2 E^\pm_\mathbf{k}/\partial k_\alpha\partial k_\beta$ is the inverse effective mass, and
$H^{\alpha\beta}_\mathbf{k} = [\Delta/(E^+_\mathbf{k} - E^-_\mathbf{k})] \,
\partial^2 (\epsilon_{\mathbf{k}+\mathbf{Q}} - \epsilon_\mathbf{k})/
\partial k_\alpha \partial k_\beta$.
The Hall coefficient is given by
\begin{equation} \label{eq:R_H}
 R_H = \frac{\sigma^{xyz}_H}{\sigma^{xx}\sigma^{yy}} ,
\end{equation}
and the Hall number by $n_H = (|e|R_H)^{-1}$. The sign is chosen such that hole pockets contribute positively and electron pockets negatively to the Hall number.

For the single-layer compound LSCO the momentum sums in the conductivity formulas are replaced by $L^{-1} \sum_\mathbf{k} \to \int \frac{d^2\mathbf{k}}{(2\pi)^2}$ in the thermodynamic limit, and for the bilayer compound YBCO by $L^{-1} \sum_\mathbf{k} \to \frac{1}{2} \sum_{k_z=0,\pi} \int \frac{d^2\mathbf{k}}{(2\pi)^2}$.

The ratio of interband and intraband contributions to the conductivities vanishes as $\Gamma^2$ in the limit $\Gamma \to 0$, and for realistic values of the scattering rate in cuprates (in the regime of interest), the interband contributions are already comparatively small. \cite{Mitscherling2018} For small $\Gamma$ the intraband contributions assume the simple form known for non-interacting band electrons, with the bare dispersion replaced by the quasi-particle bands $E_\mathbf{k}^n$. \cite{Voruganti1992}
Our numerical results presented below have been obtained from the complete expressions which include interband contributions.

To take the renormalization of the quasi-particle energies into account, we replace $E_\mathbf{k}^n$ by $\bar E_\mathbf{k}^n = Z E_\mathbf{k}^n$ as described in the preceding section.
Note that the reduction of spectral weight of single-particle excitations by the $Z$-factor does not apply to the conductivities. The reduction of the quasi-particle contribution to the propagators by $Z$ is canceled by vertex corrections to the conductivities. \cite{Nozieres1964}

%
%
%

\subsection{Results}
\label{sec:transport:results}

We now show and discuss results for the longitudinal and Hall conductivities as obtained by inserting the DMFT result for the magnetic order parameter and the $Z$-factor into the expressions for the conductivities presented above. For the scattering rate we assume the doping independent value $\Gamma = 0.025t$ corresponding to an estimate for $\rm La_{1.6-x}Nd_{0.4}Sr_xCuO_4$ (Nd-LSCO) at low temperatures. \cite{Gamma}
The magnetic gap in the zero temperature results is based on a linear extrapolation of $\Delta(p)$ as shown in Fig.~\ref{fig:gap_vs_dop}.
The zero temperature limit of $\eta(p)$ and $Z(p)$ was obtained by a linear temperature extrapolation at each doping, \cite{footnote1} and a subseqent linear fit in $p$ up to the zero temperature extrapolation of $p^*$.

\begin{figure}
\includegraphics[width=0.45\textwidth]{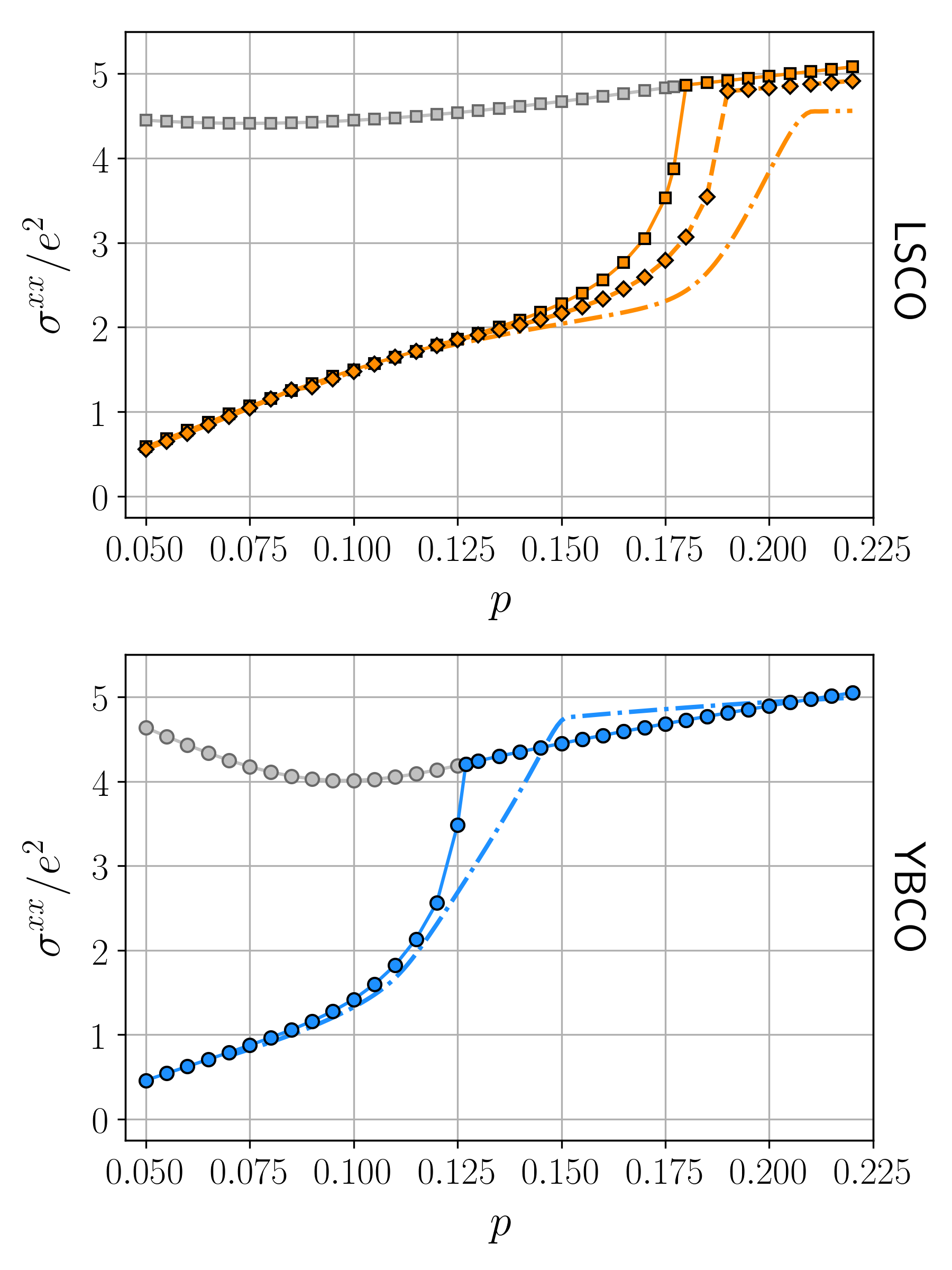}
\caption{Longitudinal conductivity as a function of doping for LSCO at $T=0.027t$ (dashed line) and $T=0.04t$ (solid line), and for YBCO at $T=0.04t$ (solid line), together with an extrapolation to zero temperature (dashed-dotted lines). The conductivity in the unstable paramagnetic phase is also shown for comparison at $T=0.04t$ (grey lines).}
\label{fig:conductivity}
\end{figure}
In Fig.~\ref{fig:conductivity} we show the longitudinal conductivity $\sigma^{xx}$ as a function of doping for LSCO parameters at $T=0.027t$ and $T=0.04t$, and for YBCO parameters at $T=0.04t$, together with an extrapolation to zero temperature. Note that $\sigma^{xx}/e^2$ is a dimensionless quantity since we use natural units where $\hbar=1$. Our results for the two-dimensional conductivity correspond to three-dimensional resistivities of the order $100\mu\Omega cm$, in agreement with experimental values. \cite{footnote2}
The expected drop below $p^*$ is clearly visible. It is particularly steep at $T > 0$, which is due to the square root type onset of the order parameter at finite temperature, see Fig.~\ref{fig:gap_vs_dop}. Since the scattering rate is fixed in our calculations, the drop of $\sigma^{xx}$ is exclusively due to a drop of charge carrier concentration related to the Fermi surface reconstruction by the magnetic gap.

\begin{figure}
\includegraphics[width=0.45\textwidth]{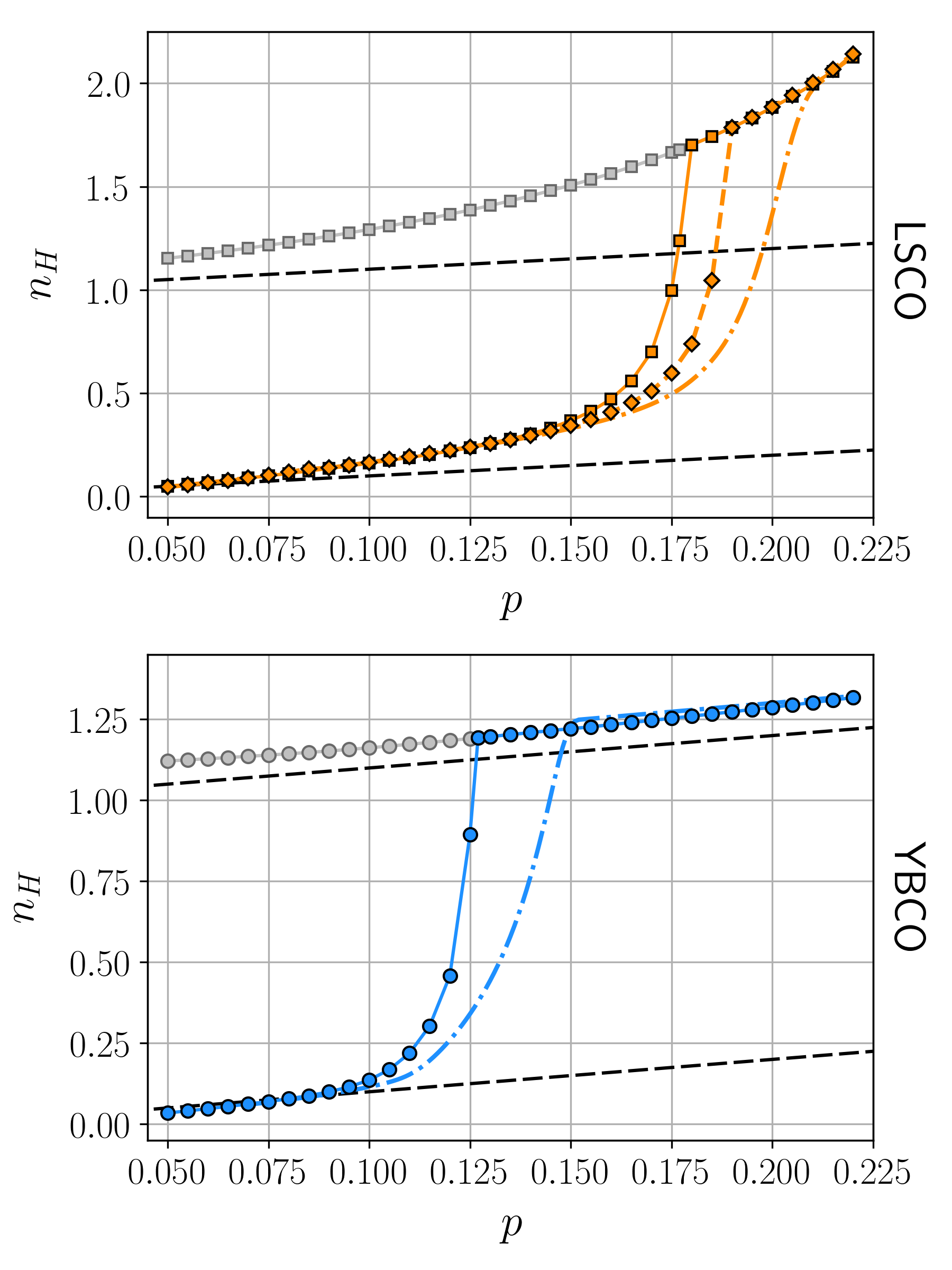}
\caption{Hall number as a function of doping for LSCO at $T=0.027t$ (dashed line) and $T=0.04t$ (solid line), and for YBCO at $T=0.04t$ (solid line), together with an extrapolation to zero temperature (dashed-dotted lines). The Hall number in the unstable paramagnetic phase is also shown for comparison at $T=0.04t$ (grey lines).
The black dashed lines correspond to the naive expectations $n_H = p$ for hole pockets
and $n_H = 1 + p$ for a large hole-like Fermi surface.}
\label{fig:hallnumber}
\end{figure}
The Hall number is plotted as a function of doping in Fig.~\ref{fig:hallnumber}, again for LSCO at $T=0.027t$ and $T=0.04t$, and for YBCO parameters at $T=0.04t$, together with an extrapolation to zero temperature. A pronounced drop is seen for doping values below $p^*$, indicating once again a drop of the charge carrier concentration.

In the high-field limit $\omega_c \tau \gg 1$ the Hall number would be exactly equal to the charge carrier density enclosed by the Fermi lines, that is, $1+p$ in the paramagnetic phase and $p$ in the magnetically ordered phase, even in the presence of electron pockets. \cite{Ashcroft1976} However, the experiments which motivated our analysis are in the {\em low-field}\/ limit $\omega_c \tau \ll 1$, since $\tau$ is rather small, and our expression for the Hall conductivity has been derived in this limit. In the low field limit the Hall number is equal to the carrier density only for a parabolic dispersion.
For low doping the Hall number $n_H(p)$ shown in Fig.~\ref{fig:hallnumber} indeed approaches the value $p$, which indicates a near-parabolic dispersion of the holes in the hole pockets for small $p$.
For large doping, in the paramagnetic phase, the Hall number is only slightly above the naively expected value $1+p$ in YBCO, while it shoots up to significantly higher values in LSCO, indicating that the dispersion of charge carriers near the Fermi surface cannot be approximated by a parabolic form. The increase of $n_H(p)$ way above $1+p$ is a precursor of a divergence at the doping $p = 0.33$, well above the van Hove point at $p=0.23$, which is due to a cancellation of positive (hole-like) and negative (electron-like) contributions to the Hall coefficient $R_H$.

\begin{figure}
\includegraphics[width=0.48\textwidth]{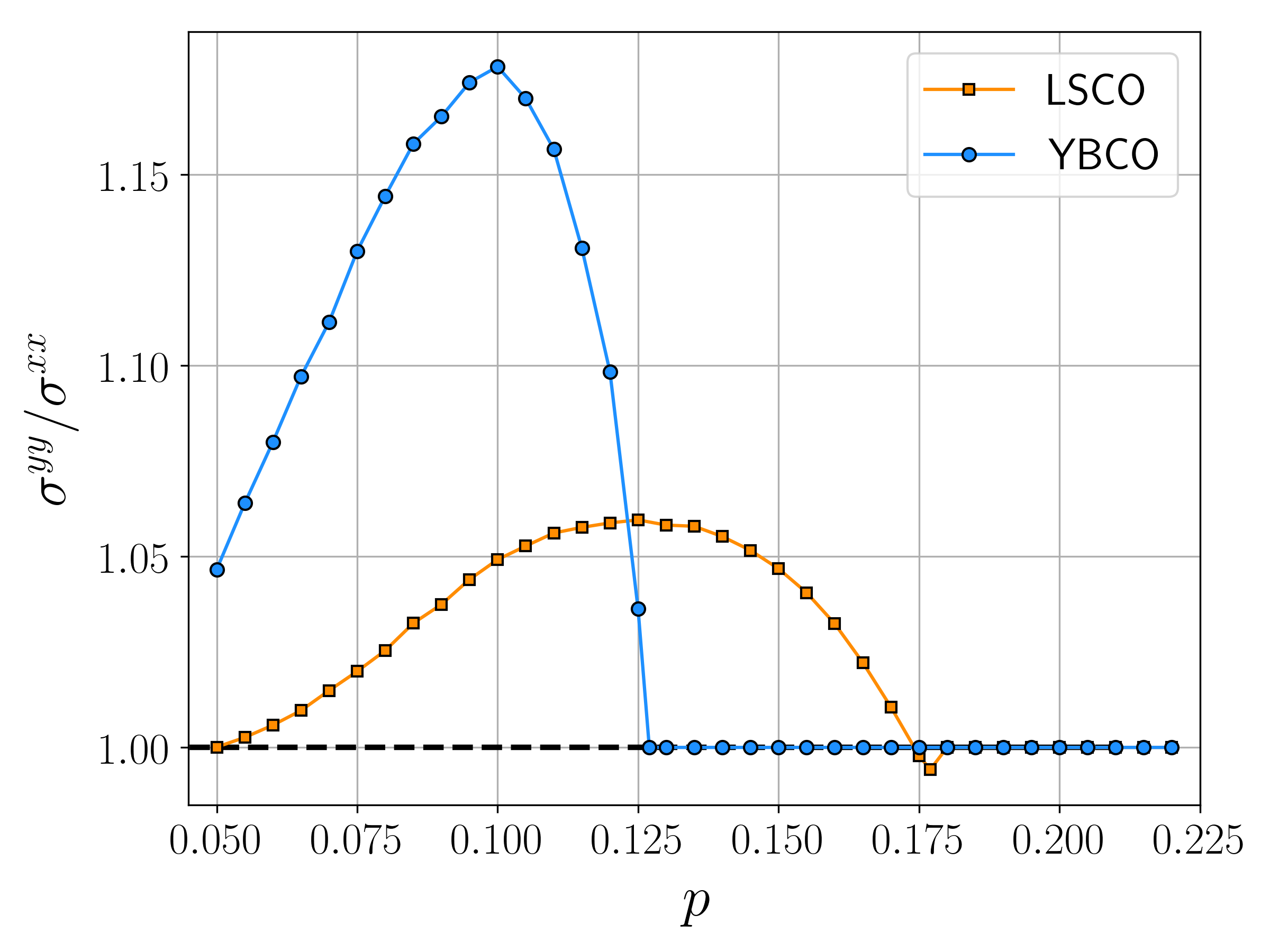}
\caption{Ratio $\sigma^{yy}/\sigma^{xx}$ as a function of doping for LSCO (orange) and YBCO (blue) at $T=0.04t$.}
\label{fig:nematicity}
\end{figure}
In Fig.~\ref{fig:nematicity} we show the ratio $\sigma^{yy}/\sigma^{xx}$ as a function of doping for LSCO and YBCO at $T = 0.04t$. The breaking of the tetragonal symmetry of the square lattice by a spiral order with $\eta > 0$ naturally leads to an anisotropy (or ``nematicity'') in the longitudinal conductivity. For a wave vector of the form $\mathbf{Q} = (\pi - 2\pi\eta,\pi)$ the conductivity in $y$-direction is somewhat larger than the one in $x$-direction. Lowering the doping below $p^*$ the anisotropy first increases due to the increasing magnetic gap, but then decreases again due to the decreasing incommensurability $\eta$ for low doping.


\subsection{Comparison to experiments}
\label{sec:transport:comparison}

Qualitatively, the pronounced drop of the longitudinal conductivity and the Hall number observed in experiment is captured by our theory.
The onset of the drop at $p^* = 0.21$ in our zero temperature extrapolation for LSCO is slightly above the experimental value $0.18$ for LSCO, \cite{Laliberte2016} but below the value $0.23$ observed for Nd-LSCO. \cite{Collignon2017} Why the observed $p^*$ differs so much between LSCO and Nd-LSCO is unclear.
For YBCO we obtain $p^* = 0.15$ while the charge carrier drop seen in experiment starts at $p^* = 0.19$. \cite{Badoux2016}
Cluster extensions of the DMFT \cite{Maier2005} yield critical doping strengths for the onset of pseudogap behavior which are also below the experimental value. \cite{Wu2018,Reymbaut2019}
Hence, for a better quantitative agreement one probably needs to go beyond the single-band Hubbard model.

The relatively narrow doping range of a few percent in which the steep charge carrier drop occurs also agrees between theory and experiment. However, the Hall number obtained from our calculations reaches the value $n_H(p) = p$ only at much lower dopings than in the experiments. The convergence is particularly slow for LSCO and can be attributed to the non-parabolic dispersion of the charge carriers in the hole-pockets. In the experiments, the behavior $n_H \approx p$ is observed over an {\em extended}\/ doping range only at low doping far below $p^*$, too. At larger doping, a few percent below $p^*$, the Hall number becomes equal to $p$ only at a {\em single}\/ doping value. Upon further reducing the doping it drops below $p$ and even becomes negative, presumably due to charge density wave order. \cite{Leboeuf2007} To obtain the steep drop of the Hall number down to $n_H = p$ and below in a theoretical description, one therefore needs to take the charge order into account. Charge order on top of spiral order was discussed by Eberlein et al. \cite{Eberlein2016}, but the corresponding transport properties were not yet computed.

For dopings $p \geq p^*$ the Hall number for YBCO is close to $1+p$ as naively expected. More precisely, it is slightly larger in agreement with the experimental observations. \cite{Badoux2016} By contrast, for LSCO parameters $n_H$ is much larger than $1+p$ for $p \geq p^*$ with an increasing deviation for larger doping. From a theoretical point of view this behavior is not surprising, since there is no reason why $n_H$ should be close to the charge carrier density for the strongly non-parabolic dispersion near van Hove filling. In the experiments $p^*$ practically coincides with van Hove filling in Nd-LSCO, and $n_H$ is nevertheless only slightly above $1+p$ for $p$ near $p^*$. \cite{Collignon2017}

A drop by a factor $p/(1+p)$ in a narrow doping range below $p^*$ has also been observed for the longitudinal conductivity in Nd-LSCO. \cite{Collignon2017} This drop corresponds to the expectation based on a Drude formula for the conductivity only if the scattering rate and the effective electron mass remain constant while the charge carrier concentration drops from $1+p$ to $p$. The drop of $\sigma^{xx}$ below $p^*$ obtained from our calculation for LSCO parameters is less pronounced (see Fig.~\ref{fig:conductivity}). Since we assumed a doping-independent scattering rate, this reduction of the drop compared to the carrier concentration ratio must be due to an increase of the average Fermi velocities below $p^*$, that is, a decrease of the effective electron mass in a Drude picture. A priori there is no reason why these quantities should remain constant when the Fermi surface gets fractionalized in pockets. Hence, the drop of the longitudinal conductivity by a factor $p/(1+p)$ observed for Nd-LSCO is probably just an accident. Of course, it can be reconciled with the theory by assuming a suitable doping-dependent enhancement of the scattering rate below $p^*$. \cite{Storey2017,Chatterjee2017}

A pronounced temperature and doping-dependent in-plane anisotropy (nematicity) of the longitudinal conductivities has been observed in YBCO by Ando {\em et al.} \cite{Ando2002}
The maximal conductivity ratios $\sigma^{yy}/\sigma^{xx}$ observed in these experiments are much larger (up to 2.5) than those obtained in our calculation.

%
%
%
%

\section{Conclusion}

Using the two-dimensional Hubbard model to describe the valence electrons in the $\rm CuO_2$-planes of high-$T_c$ cuprates, we have shown that antiferromagnetic spin-density wave order can explain the pronounced drop of the charge carrier density at the onset of the pseudogap regime observed in recent transport experiments in high magnetic fields. \cite{Badoux2016,Laliberte2016,Collignon2017,Proust2019}

The amplitude and the wave vector of the spin-density wave was computed for a strong Hubbard interaction by DMFT. The wave vector has the form $\mathbf{Q} = (\pi-2\pi\eta,\pi)$, where the incommensurability $\eta$ increases with doping. The magnetic gap $\Delta$ decreases monotonically as a function of doping and vanishes at a critical doping $p^*$. An extrapolation of the numerical results (obtained at low finite $T$) to zero temperature yields an approximately linear doping dependence $\Delta(p) \propto p^*-p$ in a broad doping range below $p^*$.
The magnetic order leads to a Fermi surface reconstruction with electron and hole pockets. Electron pockets exist only in a restricted doping range near $p^*$. Due to a strong momentum dependence of the spectral weight along the reconstructed Fermi surface, the spectral function for single-particle excitations seen in angular resolved photoemission seems to exhibit Fermi arcs instead of hole-pockets. The backbending edges of the arcs in our results could be suppressed by taking a more realistic momentum-dependent scattering rate into account.

Longitudinal and Hall conductivities were computed by inserting the magnetic gap, the magnetic wave vector, and the quasi-particle renormalization $Z$ as obtained from the DMFT into transport equations for spin-density wave states with a phenomenological scattering rate. \cite{Mitscherling2018} A pronounced drop of the longitudinal conductivity and the Hall number in a narrow doping range of few percent below $p^*$ is obtained in agreement with the corresponding high-field experiments. The doping range in which electron pockets exist matches approximately with the range of the steepest Hall number drop, but there is no simple theoretical relation between these two features.
For $p>p^*$ the calculated Hall number $n_H(p)$ is close to the naively expected value $1+p$ for YBCO parameters, but significantly higher for LSCO. From a theoretical point of view this is not surprising since the band structure near the Fermi surface of LSCO cannot be approximated by a parabolic band in a broad doping range around $p^*$.
For $p<p^*$ the Hall number approaches the value $p$ from above, but converges to this limiting value only for dopings well below the point where the electron pockets disappear. $n_H(p) \approx p$ is obtained only in a regime where the hole pockets are sufficiently small so that the quasi-particle dispersion in the pockets is approximately parabolic. In the cuprates $n_H(p)$ does not {\em converge}\/ to $p$ but rather {\em crosses}\/ the line $n_H(p) = p$ at a doping value a few percent below $p^*$, and becomes negative at lower doping, presumably due to electron pockets associated with charge density wave order. \cite{Leboeuf2007} Computing charge transport properties in the presence of charge density wave order on top of magnetic order could thus be an interesting extension of our work.

The zero temperature extrapolation of our results for the magnetic order parameter as a function of doping yields $p^* = 0.21$ for LSCO parameters and $p^* = 0.15$ for YBCO. These values are in the correct range, but we are obviously not in a position to provide accurate predictions for the experimentally observed critical doping. For a better agreement one probably needs a material dependent modelling beyond the single-band Hubbard model.


\begin{acknowledgments}
We are grateful to A.~Greco, G.~Grissonnanche, S.~Kivelson, T.~Maier, L.~Taillefer, A.-M.~Tremblay, and S.~Verret for valuable discussions.
\end{acknowledgments}



\begin{thebibliography}{2}

\bibitem{Broun2008} D.~M.~Broun, What lies beneath the dome?,
 Nat. Phys. {\bf 4}, 170 (2008).

\bibitem{Badoux2016} S.~Badoux, W.~Tabis, F.~Lalibert\'e, G.~Grissonnanche,
 B.~Vignolle, D.~Vignolles, J.~B\'eard, D.~A.~Bonn, W.~N.~Hardy, R.~Liang,
 N.~Doiron-Leyraud, L.~Taillefer, and C.~Proust, Change of carrier density
 at the pseudogap critical point of a cuprate superconductor,
 Nature (London) {\bf 531}, 210 (2016).

\bibitem{Laliberte2016} F.~Lalibert\'e, W.~Tabis, S.~Badoux, B.~Vignolle, N.~Momono,
 T.~Kurosawa, K.~Yamada, H.~Takagi, N.~Doiron-Leyraud, C.~Proust, and L.~Taillefer,
 Origin of the metal-to-insulator crossover in cuprate superconductors,
 arXiv:1606.04491 (2016).
 
\bibitem{Collignon2017} C.~Collignon, S.~Badoux, S.~A.~A.~Afshar, B.~Michon,
 F.~Lalibert\'e, O.~Cyr-Choini\`ere, J.-S.~Zhou, S.~Licciardello, S.~Wiedmann,
 N.~Doiron-Leyraud, and L.~Taillefer, Fermi-surface transformation across
 the pseudogap critical point of the cuprate superconductor
 $\rm La_{1.6-x} Nd_{0.4} Sr_x CuO_4$, Phys. Rev. B {\bf 95}, 224517 (2017).

\bibitem{Proust2019} C.~Proust and L.~Taillefer,
 The Remarkable Underlying Ground State of Cuprate Superconductors,
 Annu. Rev. Condens. Matter Phys. {\bf 10}, 409 (2019).
 
\bibitem{Frachet2019} M.~Frachet, I.~Vinograd, R.~Zhou, et al.,
 Hidden magnetism at the pseudogap critical point of a high temperature superconductor,
 arXiv 1909.10258 (2019).

\bibitem{Keimer2015} B.~Keimer, S.~A.~Kivelson, M.~R.~Norman, S.~Uchida, and J.~Zaanen,
 From quantum matter to high-temperature superconductivity in copper oxides,
 Nature {\bf 518}, 179 (2015).
 
\bibitem{Storey2016} J.~G.~Storey,
 Hall effect and Fermi surface reconstruction via electron pockets in the high
 $T_c$ cuprates, Europhys. Lett. {\bf 113}, 27003 (2016).
 
\bibitem{Storey2017} J.~G.~Storey,
 Simultaneous drop in mean free path and carrier density at the pseudogap onset
 in high-$T_c$ cuprates,
 Supercond. Sci. Technol. {\bf 30}, 104008 (2017).
 
\bibitem{Verret2017} S.~Verret, O.~Simard, M.~Charlebois, D.~S\'en\'echal, and A.~M.~S.~Tremblay,
 Phenomenological theories of the low-temperature pseudogap: Hall number, specific heat, and Seebeck coefficient, Phys. Rev. B {\bf 96}, 125139 (2017).
 
\bibitem{Eberlein2016} A.~Eberlein, W.~Metzner, S.~Sachdev, and H.~Yamase,
 Fermi Surface Reconstruction and Drop in the Hall number due to Spiral
 Antiferromagnetism in High-$T_c$ Cuprates, Phys. Rev. Lett. {\bf 117}, 187001 (2016).
 
\bibitem{Chatterjee2017} S.~Chatterjee, S.~Sachdev, and A.~Eberlein,
 Thermal and electrical transport in metals and superconductors across
 antiferromagnetic and topological quantum transitions,
 Phys. Rev. B {\bf 96}, 075103 (2017).

\bibitem{Mitscherling2018} J.~Mitscherling and W.~Metzner,
 Longitudinal conductivity and Hall coefficient in two-dimensional metals with
 spiral magnetic order,
 Phys. Rev. B {\bf 98}, 195126 (2018).
 
\bibitem{Caprara2017} S.~Caprara, C.~Di Castro, G.~Seibold, and M.~Grilli,
 Dynamical charge density waves rule the phase diagram of cuprates,
 Phys. Rev. B {\bf 95}, 224511 (2017).
 
\bibitem{Sharma2018} G.~Sharma, S.~Nandy, A.~Taraphder, and S.~Tewari,
 Suppression of the Hall number due to charge density wave order in high-$T_c$ cuprates,
 Phys. Rev. B {\bf 97}, 195153 (2018).
 
\bibitem{Maharaj2017} A.~V.~Maharaj, I.~Esterlis, Y.~Zhang, B.~J.~Ramshaw, and S.~A.~Kivelson,
 Hall number across a van Hove singularity,
 Phys. Rev. B {\bf 96}, 045132 (2017).
 
\bibitem{Qi2010} Y.~Qi and S.~Sachdev,
 Effective theory of Fermi pockets in fluctuating antiferromagnets,
 Phys. Rev. B {\bf 81}, 115129 (2010).
 
\bibitem{Chatterjee2016} S.~Chatterjee and S.~Sachdev,
 Fractionalized Fermi liquid with bosonic chargons as a candidate for the
 pseudogap metal, Phys. Rev. B {\bf 94}, 205117 (2016).
 
\bibitem{Scheurer2018} M.~S.~Scheurer, S.~Chatterjee, W.~Wu, M.~Ferrero, A.~Georges,
 and S.~Sachdev, Topological order in the pseudogap metal,
 Proc. Natl. Acad. Sci. U.S.A. {\bf 115}, E3665 (2018).
 
\bibitem{Morice2017} C.~Morice, X.~Montiel, and C.~P\'epin,
 Evolution of Hall resistivity and spectral function with doping in the SU(2) theory of cuprates,
 Phys. Rev. B {\bf 96}, 134511 (2017).
 
\bibitem{Yang2006} K.-Y.~Yang, T.~M.~Rice, and F.-C.~Zhang,
 Phenomenological theory of the pseudogap state,
 Phys. Rev. B {\bf 73}, 174501 (2006).
 
\bibitem{Charlebois2017} M.~Charlebois, S.~Verret, A.~Foley, O.~Simard,
 D.~S\'en\'echal, and A.-M.~S.~Tremblay,
 Hall effect in cuprates wit an incommensurate collinear spin-density wave,
 Phys. Rev. B {\bf 96}, 205132 (2017).

\bibitem{Scalapino2012} D.~J.~Scalapino,
 A common thread: The pairing interaction for unconventional superconductors,
 Rev. Mod. Phys. {\bf 84}, 1383 (2012).
 
\bibitem{Yamase2016} H.~Yamase, A.~Eberlein, and W.~Metzner,
 Coexistence of Incommensurate Magnetism and Superconductivity in the
 Two-Dimensional Hubbard Model, Phys. Rev. Lett. {\bf 116}, 096402 (2016).
 
\bibitem{Metzner1989a} W.~Metzner,
 Variational theory for correlated lattice fermions in high dimensions,
 Z. Phys. B {\bf 77}, 253 (1989).
 
\bibitem{Fleck1999} M. Fleck, A.I. Lichtenstein, A.M. Ole\'{s} and L. Hedin,
 Spectral and transport properties of doped Mott-Hubbard systems with incommensurate
 magnetic order,
 Phys. Rev. B {\bf 60}, 5224 (1999).

\bibitem{Vilardi2018} D.~Vilardi, C.~Taranto, and W.~Metzner,
 Dynamically enhanced magnetic incommensurability: Effects of local dynamics on
 non-local spin-correlations in a strongly correlated metal,
 Phys. Rev. B {\bf 97}, 235110 (2018).
 
\bibitem{Montorsi1992} A.~Montorsi,
 \emph{The Hubbard Model: A Reprint Volume} (World Scientific, 1992). 

\bibitem{Anderson1987} P.W.~Anderson,
The Resonating Valence Bond State in $\rm La_2CuO_4$ and Superconductivity,
Science {\bf 235}, 1196 (1987).
 
\bibitem{Andersen1995} O.~K.~Andersen, A.I.~Liechtenstein, O.~Jepsen, and F.~Paulsen, 
 LDA energy bands, low-energy Hamiltonians, $t'$, $t''$, $t_\bot(\mathbf{k})$,
 and $J_\bot$, J. Phys. Chem.  Solids \textbf{56}, 1573 (1995).

\bibitem{Pavarini2001} E.~Pavarini, I.~Dasgupta, T.~Saha-Dasgupta, O.~Jepsen,
 and O.~K.~Andersen,
 Band-Structure Trend in Hole-Doped Cuprates and Correlation with $T_{\rm cmax}$,
 Phys. Rev. Lett. \textbf{87}, 047003 (2001).
 
\bibitem{Fresard1991} R.~Fresard, M.~Dzierzawa, and P.~W\"olfle,
 Slave-Boson Approach to Spiral Magnetic Order in the Hubbard Model,
 Europhys. Lett. \textbf{15}, 325 (1991).
 
\bibitem{Igoshev2010} 
 P.~A.~Igoshev, M.~A.~Timirgazin, A.~A.~Katanin, A.~K.~Arzhnikov, and V.~Yu.~Irkhin,
 Incommensurate magnetic order and phase separation in the two-dimensional Hubbard model
 with nearest- and next-nearest-neighbor hopping, 
 Phys. Rev. B \textbf{81}, 094407 (2010).
 
\bibitem{Shraiman1989} B.~I.~Shraiman and E.~D.~Siggia,
 piral phase of a doped quantum antiferromagnet,
 Phys. Rev. Lett. \textbf{62}, 1564 (1989).
  
\bibitem{Kotov2004} V.~N.~Kotov and O.~P.~Sushkov,
 Stability of the spiral phase in the two-dimensional extended t-J model,
 Phys. Rev. B \textbf{70}, 195105 (2004).
  
\bibitem{Georges1996} A.~Georges, G.~Kotliar, W.~Krauth, and M.~J.~Rozenberg, 
 Dynamical mean-field theory of strongly correlated fermion systems
 and the limit of infinite dimensions,
 Rev. Mod. Phys. \textbf{68}, 13-125 (1996).
 
\bibitem{Metzner1989} W.~Metzner and D.~Vollhardt,
 Correlated Lattice Fermions in $d=\infty$ Dimensions,
 Phys. Rev. Lett. \textbf{62}, 1066 (1989).
 
\bibitem{Georges1992} A.~Georges and G.~Kotliar,
 Hubbard Model in Infinite Dimensions,
 Phys. Rev. B \textbf{45}, 6479 (1992).

\bibitem{Goto2016} S. Goto, S. Kurihara and D. Yamamoto,
 Incommensurate spiral magnetic order on anisotropic triangular lattice:
 Dynamical mean-field study in a spin-rotating frame,
 Phys. Rev. B, \textbf{94}, 245145 (2016). 
 
\bibitem{Kitatani2018} M.~Kitatani, T.~Sch\"afer, H.~Aoki, and K.~Held,
 Why T$_c$ is so low in high-T$_c$ cuprates: importance of the dynamical
 vertex structure, Phys. Rev. B \textbf{99}, 041115 (2019).
 
\bibitem{Parcollet2015} O.~Parcollet, M.~Ferrero, T.~Ayral, H.~Hafermann,
 I.~Krivenko, L.~Messio, and P.~Seth, 
 TRIQS: A Toolbox for Research on Interacting Quantum Systems, 
 Comp. Phys. Comm. \textbf{196}, 398 (2015)
 
\bibitem{Schulz1990} H.~J.~Schulz,
 Incommensurate Antiferromagnetism in the Two-Dimensional Hubbard-Model,
 Phys. Rev. Lett. {\bf 64}, 1445 (1990).

\bibitem{Yamada1998} K.~Yamada, C.~H.~Lee, K.~Kurahashi, et al.,
 Doping dependence of the spatially modulated dynamical spin correlations and the
 superconducting transition temperature in $\rm La_{2-x} Sr_x Cu O_4$,
 Phys. Rev. B {\bf 57}, 6165 (1998).
 
\bibitem{Haug2010} D.~Haug, V.~Hinkov, Y.~Sidis, N.~B.~Christensen, A.~Ivanov,
 T.~Keller, C.~T.~Lin, and B.~Keimer, Neutron scattering study of the magnetic
 phase diagram of underdoped $\rm YBa_2Cu_3O_{6+x}$,
 New J. Phys. {\bf 12}, 105006 (2010).
  
\bibitem{Voruganti1992} P.~Voruganti, A.~Golubentsev, and S.~John,
 Conductivity and Hall effect in the two-dimensional Hubbard model,
 Phys.\ Rev.\ B {\bf 45}, 13945 (1992).
  
\bibitem{Nozieres1964} P.~Nozi\`eres,
 \emph{Theory of Interacting Fermi Systems} (Benjamin, Amsterdam, 1964).
  
\bibitem{Gamma} Ref.~\onlinecite{Mitscherling2018}, Sec.~III, and L.~Taillefer,
 private communication.

\bibitem{footnote1} The linear extrapolation in temperature is based on data for $T=0.027t$
and $T=0.04t$ for LSCO, and data for $T=0.04t$ and $T=0.05t$ for YBCO.
 
\bibitem{footnote2} $h/e^2$ is the von Klitzing constant $R_K \approx 25813\Omega$. The two-dimensional conductivity of a CuO-layer in SI-units is thus obtained by multiplying our dimensionless result by $2\pi/25813\Omega$. To obtain the conductivity of the three-dimensional sample, one has to divide by the average distance between the layers.
 
\bibitem{Ashcroft1976} N.~W.~Ashcroft and N.~D.~Mermin,
 \emph{Solid State Physics} (Saunders College, Philadelphia, 1976).

\bibitem{Maier2005} T.~Maier, M.~Jarrell, T.~Pruschke, and M.H.~Hettler, 
 Quantum cluster theories, Rev. Mod. Phys. \textbf{77}, 1027 (2005).

\bibitem{Wu2018} W.~Wu, M.~Scheurer, S.~Chatterjee, S.~Sachdev, A.~Georges,
 and M.~Ferrero,
 Pseudogap and Fermi-Surface Topology in the Two-Dimensional Hubbard Model,
 Phys. Rev. X {\bf 8}, 021048 (2018).
 
\bibitem{Reymbaut2019} A.~Reymbaut, S.~Bergeron, R.~Garioud, M.~Th\'enault,
 M.~Charlebois, P.~S\'emon, and A.-M.~S.~Tremblay,
 Pseudogap, van Hove Singularity, Maximum in Entropy and Specific Heat for
 Hole-Doped Mott Insulators, ArXiv:1905.02326 (2019).

\bibitem{Leboeuf2007} D.~LeBoeuf, N.~Doiron-Leyraud, J.~Levallois, R.~Daou,
 J.-B.~Bonnemaison, N.~E.~Hussey, L. Balicas, B.~J.~Ramshaw, R.~Liang, D.~A.~Bonn,
 W.~N.~Hardy, S.~Adachi, C.~Proust, and L.~Taillefer,
 Electron pockets in the Fermi surface of hole-doped high-$T_c$ superconductors,
 Nature {\bf 450}, 533 (2007).

\bibitem{Ando2002} Y.~Ando, K.~Segawa, S.~Komiya, and A.~N.~Lavrov,
 Electric Resistivity Anisotropy from Self-Organized One Dimensionality in
 High-Temperature Superconductors,
 Phys. Rev. Lett. {\bf 88}, 137005 (2002).

\end{thebibliography}
\end{document}